\documentclass[article,11pt]{amsart}
\usepackage{amsaddr}
\usepackage{amssymb, amsthm, amsfonts, amsgen, stmaryrd}
\usepackage{slashed}
\usepackage[english]{babel}
\usepackage{fullpage}
\usepackage[centertags]{amsmath}
\usepackage{indentfirst}
\usepackage{fancyhdr}
\usepackage[dvips]{graphicx}
\usepackage{psfrag}
\usepackage{enumerate}
\numberwithin{equation}{section}
\usepackage[latin1]{inputenc}

\usepackage{hyperref}
\hypersetup{
pdftitle={},%
pdfauthor={},%
pdfsubject={},%
pdfkeywords={},%
colorlinks=true,%
linkcolor=blue,%
citecolor=red,%
linktocpage=true,%
hyperfootnotes=true,%
pageanchor=true
}
\usepackage{color}

\title{K\"ahler geometry on complex projective spaces via reduction and unfolding} 
\date{9 september  2018}
\author{Giuseppe Marmo,  Alessandro Zampini}
\address{Dipartimento di Fisica ``E. Pancini'', Universit\`a di Napoli Federico II, \\
Via Cintia - 80126 Napoli, Italy
\\
INFN - Sezione di Napoli, Via Cintia - 80126 Napoli, Italy } 
\email{marmo@na.infn.it}
 \email{azampini@na.infn.it}

\newcommand{\nn}{\nonumber}

\newcommand{\dd}{{\rm d}}

\newcommand{\figureheight}{8cm}
\newcommand{\putfig}[2]{\begin{figure}[htp]
        \special{isoscale c:/itex/texfig/#1.wmf, \the\hsize \figureheight}
        \vspace{\figureheight}
        \caption{#2}\label{fig:#1}
        \end{figure}}
\newcommand{\pictureheight}{4cm}
\newcommand{\putpicture}[2]{\begin{figure}[htp]
        \special{isoscale c:/itex/texfig/#1.wmf, \the\hsize \pictureheight}
        \vspace{\pictureheight}
        \caption{#2}\label{fig:#1}
        \end{figure}}

\newcommand{\beqa}{\begin{eqnarray}}
\newcommand{\eeqa}{\end{eqnarray}}
\newcommand{\beq}{\begin{equation}}
\newcommand{\eeq}{\end{equation}}

\newcommand{\del}{\partial}

 \font\mybb=msbm10 at 12pt
\def\bb#1{\hbox{\mybb#1}}

\newcommand{\e}{\mathrm{e}}

\newcommand{\R}{{\mathbb{R}}}

\newcommand{\C}{\mathbb{C}}

\newcommand{\Hi}{\mathcal H}
\newcommand{\PP}{\mathbb P}
\newcommand{\dequ}{\frac{\del}{\del q_1}}
\newcommand{\deqd}{\frac{\del}{\del q_2}}
\newcommand{\depu}{\frac{\del}{\del p_1}}
\newcommand{\depd}{\frac{\del}{\del p_2}}

\newcommand{\mfu}{\mathfrak{u}}

\begin{document}

\thispagestyle{empty}

\begin{abstract}
We review how a reduction procedure along a principal fibration and an unfolding procedure associated to a suitable momentum map allows to describe the K\"ahler geometry of  a finite dimensional complex projective space.

\end{abstract}


\maketitle
\tableofcontents

\section{Introduction}
\label{sec:intro}

Any picture (i.e. a mathematical formulation) for the dynamics of a physical system requires to identify  a convex set -- denote it by $\mathcal S$ --  of \emph{states}, which represent the maximal information about the system, together with a real vector space -- denote it by $\mathcal O$ -- of \emph{observables} (i.e. measurable quantities) for the system. These sets are paired, that is there exists a map, called  \emph{pairing},
$$
\mu\,:\,\mathcal O\,\times\,\mathcal S\quad\to\quad\mathcal P,
$$
with $\mathcal P$ the set of probability measures on the real line $\R$. Given a state $\rho\,\in\,\mathcal S$ and an observable $A\,\in\,\mathcal O$, the quantity $\mu(A,\rho)(\Delta)$ provides the probability that the measurement of $A$ while the system is in the state $\rho$ gives a result in $\Delta$, with $\Delta$ an element in the Borel $\sigma$-algebra over $\R$. The time \emph{evolution} of a physical system with such $(\mathcal O, \mathcal S, \mu)$ is described by a one parameter group $\Phi_t$ (being $t$ the time parameter) of automorphisms defined either on the space of observables or on the space of states or on the space of probability measures. Basic requirements for the description of a physical system end with  a rule to describe \emph{composite} systems.  

A geometric formulation of classical mechanics is based on the notion of a differentiable manifold $M$: points $m\in M$ give the pure states of the system, real valued (smooth) functions defined on $M$ give the observables. The pairing between them is given by the evaluation of a function $f$ on $m$: the real value $f(m)$ provides the result of the measurement of (the observable) $f$ when the (pure) state of the system is $m$. The time evolution of the system is given by a one parameter  flow on $M$ whose infinitesimal generator is a vector field. States which are not pure, also called densities,  are described by positive measures $\dd \mu=\rho(m)\dd m $ where $\rho$ (the Radon-Nikodym derivative with respect to the Lebesgue measure $\dd m$) is not negative and normalised (i.e. $\int_M\dd\mu=1$).  On a density state, the evaluation is replaced by the average $\langle f\rangle_{\rho}=\int_M\dd\mu \,f$. 

The interpretation of the pairing as a duality can be algebraically  described by recalling  that a state for a unital  $C^*$-algebra $\mathcal A$ with Banach dual $\mathcal A'$, is an element $\rho\in\mathcal A'$ which is positive and normalised.  If $\mathcal A=C(M)$ is the commutative $C^*$-algebra of continuous functions on a compact  Hausdorff space $M$ (whose selfadjoint elements represent the observables of the system),  then its state space $\mathcal S(\mathcal A)$ consists of all probability measures  on $M$. The set $\mathcal S(\mathcal A)$ is a compact convex subset of $\mathcal A'$ (equipped with the weak $^*$-topology), its extremal points (i.e. pure states) are identified with points $m$ (i.e. $\delta$-like measures on $M$).

Within the Dirac's and Schr\"odinger's picture of quantum mechanics each physical system is associated to a separable Hilbert space, say $\mathcal H$, and states $\mathcal S$ are given by density operators on $\mathcal H$ (notice that a linear structure over $\C$ allows indeed to describe interference phenomena). Observables are  given by linear self-adjoint operators on $\mathcal H$, and the Born's  interpretation reads  
$$
\mu(A,\rho)(\Delta)\,=\,\rm{Tr}(\rho\,E_A(\Delta)),
$$
where $\rho$ is a density operator in $\mathcal S$, $A$ is the self-adjoint operator describing an observable, $E_A(\Delta)$ is the projector in $\mathcal H$ coming from the spectral resolution\footnote{Adopting the Dirac's bra-ket notation, if $A=A^{\dagger}$ has a part of point spectrum $\sigma_P(A)$ with $A\mid\e_k\rangle=\lambda_k\mid e_k\rangle$ and a part of continuous spectrum $\sigma_C(A)$ with  $A\mid \varphi_a\rangle=a\mid \varphi_a\rangle$, then there is a spectral resolution 
$$1\,=\,\sum_{\lambda_k\,\in\,\sigma_P(A)}\mid e_k\rangle\langle e_k\,+\,\int_{\sigma_C(A)}\dd a \mid\varphi_a\rangle\langle \varphi_a\mid$$ on $\mathcal H$, so that $$E_A(\Delta)\,=\,\sum_{k:\lambda_k\in\Delta}\mid e_k\rangle\langle e_k\mid\,+\,\int_{\Delta}\dd a\mid\varphi_a\rangle\langle\varphi_a\mid.$$} of $A$ for any Borel set $\Delta\subset\R$. 
The evolution is given by a one parameter group $U_t$ of unitary operators on $\mathcal H$, whose infinitesimal generator satisfies the Schr\"odinger equation 
$$
i\hbar\frac{\del\psi}{\del t}\,=\,H\psi
$$
with $\psi\,\in\,\mathcal H$ and $H$ a self-adjoint operator on $\mathcal H$  which is usually required to be bounded from below. 
When two systems with associated Hilbert spaces  $\mathcal H_1$ and $\mathcal H_2$ are composed, the  Hilbert space corresponding to  the composition is given by the tensor product $\mathcal H_{12}=\mathcal H_1\otimes\mathcal H_2$. Notice that the existence of pure states for $\mathcal H_{12}$ which are not separable, i.e. can not be written as the tensor product of a pure state on $\mathcal H_1$ times a pure states on $\mathcal H_2$ originates the problem of the entanglement.

An alternative picture of quantum mechanics comes as a development (see \cite{segal, ha-ka}) of  Heisenberg's (Born, Jordan, von Neumann) analysis in terms of infinite dimensional matrices. One identifies the observables of a physical system  as the real (i.e. Hermitian, or self-adjoint) elements  $A=A^*$ of a non commutative $C^*$-algebra $\mathcal A$. Composing two systems amounts to consider the (suitably defined) tensor product of the individual algebras. The pairing function is again given in terms of the spectrum of an element $A$, the time evolution is formulated as the adjoint action of the unitary elements $u(t)\,\in\,\mathcal A$ with $uu^*=u^*u=1$. The infinitesimal generator for such action can be written as
$$
i\hbar\frac{\dd A}{\dd t}\,=\,[H,A] 
$$
in terms of the commutator with a self-adjoint element $H$. The relations between the  two pictures are analysed\footnote{For an interesting overview of such relations, as well as for the theory on $C^*$-algebras,  we refer the reader to the first three chapters of \cite{gianfa1}, the introduction of \cite{klaas-book}, the lecture notes \cite{klaas-notes}.}  through the G.N.S. theorem, which states that any non commutative $C^*$-algebra is isomorphic to a $^*$-subalgebra of the set $\mathcal B(\mathcal H)$ of bounded operators on a separable Hilbert space $\mathcal H$.

A natural geometric description to the notion of state for a quantum mechanical system is again given in terms of states of the $C^*$-algebra $\mathcal B(\mathcal H)$. One has that
$$
\mathcal S(\mathcal B(\mathcal H))\,=\,\{\rho=\rho^{\dagger}\,\in\,\mathcal{B}(\mathcal H)\,:\,\rho\geq 0, \,{\rm Tr}(\rho)=1\}
$$ 
gives the set of normal states.  This is the set of density operators on $\mathcal H$ (we denote it by $\mathcal D$), and it is  weakly $^*$-compact and convex. Its extremal points, the pure states, are characterised by the further condition that $\rho^2=\rho$. This means that pure states of a quantum mechanical system can be identified with rank one projectors on $\mathcal H$, i.e. with elements of the complex projective space\footnote{We denote by $\mathcal H_0$ the space $\mathcal H\backslash\{0\}$ and by $\C_{0}$ the space $\C\backslash\{0\}$.} $\mathbb P\mathcal H\,=\,\mathcal H_0/\C_{0}$.

Assume, within the Hamiltonian description for a classical dynamics, that   $(V=\R^{2N}, \omega=\dd q^a\wedge\dd p_a)$ is a canonical phase space.  A Weyl system is a unitary projective representation $D\,:\,V\,\to\,\mathcal U(\mathcal H)$  of the abelian group $(V, +)$ on a separable Hilbert space, such that
$$
D(v_1)D(v_2)D^{\dagger}(v_1)D^{\dagger}(v_2)\,=\,e^{i\omega(v_1,v_2)\hbar}.
$$
Via such a set  of so called Displacement operators one defines, on a suitable domain,  the map $W\,:\,\rm{Op}(\mathcal H)\,\to\,\mathcal F(\R^{2N})$ given (we denote by $\{z\}$ the coordinate functions  on the phase space $V=\R^{2N}$ and by $\{w\}$ their Fourier dual coordinates) as
$$
W_A(z)\,=\,\int_{\R^{2N}}\frac{\dd w}{(2\pi\hbar)^N}\, e^{-i\omega(w,z)/\hbar}\,{\rm Tr}[A\,D^{\dagger}(w)]
$$
that associates, to a suitable operator $A$ on $\mathcal H$, its \emph{Wigner symbol}, i.e. a function $W_A$ on\footnote{With $W$  proven to be injective, the non commutative Moyal algebra is recovered as the set of  Wigner symbols equipped with the product  defined by  $(W_A*W_B)(z)=W_{AB}(z)$. See \cite{marse}. } 
 the classical phase space $\R^{2N}$. In general, the Wigner symbol $W_{\rho}$ of a density operator $\rho$ on $\mathcal H$ is not a probability distribution on the classical phase space, since it can assume negative values. The notion of Radon transform for integrable functions on $M$ allows to study under which conditions (see \cite{tomo-team}) both classical and quantum states can be described as functions (\emph{tomograms}) on a suitable character space dual to the classical  space for a given quantum dynamics.  
 
Although built up in terms of linear algebraic structures, quantum mechanics can be described within the formalism of differential geometry \cite{cimm, cr}. Allowing non linear transformations, this approach has proven interesting and fruitful in studying for example problems of entanglement reduction, separability, decoherence. The aim of this paper is to review the manifold structure of the set of pure states for a finite dimensional quantum system, and in particular to show how 
the Hermitian structure on the (initial) Hilbert space $\Hi$ induces a K\"ahlerian structure on the corresponding complex projective space.  

The first method we describe is based on a reduction procedure of suitable tensors  on $\Hi$ along the fibration $\Hi_0\,\to\,\Hi_0/{\C_0}$. The second method  is based on the properties of the momentum map associated to the coadjoint action \cite{k1,k2,s1} of the unitary group on the dual of its Lie algebra. 
The space of density operators $\mathcal D$ is a convex subset in the space of selfadjoint operators on $\mathcal H$. 
For a finite dimensional Hilbert space $\mathcal H=\C^N$, the set $\mathcal D$ is   a subset of the vector space dual\footnote{This duality comes from the non degenerate canonical scalar product on the Lie algebra $\mathfrak u_N$.} $\mathfrak u^*_N$ of the Lie algebra $\mathfrak u_N$ corresponding to the unitary group ${\rm U}(N)$. It is clearly $\mathcal D\,=\,\cup_{k=1,\dots,N}\mathcal D^k$ with $\mathcal D^k$  the set of density operators of rank $k$. The coadjoint action of the group ${\rm U}(N)$ on $\mathfrak u^*_N$ meaningfully restricts to each $\mathcal D^k$. Such action is transitive on the set of pure states $\mathcal D^1$, which therefore has a canonical ($(2N-2)$-dimensional) manifold structure, while it is not transitive on $\mathcal D^k$ for $k>1$. Each ${\rm U}(N)$ orbit is identified by the common spectrum of any one of its elements. 
It turns out that the spaces $\mathcal D^k$ are smooth and connected submanifolds in $\mathfrak u^*_N$ of real dimension $(2Nk-k^2-1)$, with $\mathcal D$ being a stratified manifold, where the  stratification is indexed by the rank $k$. Each stratum can indeed be \cite{prep} considered an orbit of a non linear action of the complexification ${\rm SL}(N)$ of ${\rm SU}(N)$. 

 The literature on this subject is rich. We mention \cite{strocchi}, where the idea of studying a finite level quantum dynamics in terms of complex variables and  \cite{cirelli,cirelli1}, where the problem has been considered for infinite dimensional Hilbert spaces.  We mention \cite{alek, cg-gm, gkm, gkm1, cm, mmsz, cc-gm} and refer the reader to the bibliography in these papers. 

\medskip

{\bf Acknowledgments.} This paper originated from the talk that one of us (G.M.) delivered at the conference {\it Trails in quantum mechanics and surroundings}, held in SISSA (Trieste) in january 2018. It is a pleasure for us to dedicate this paper to Gianfausto Dell'Antonio for his 85th birthday. 

\section{K\"ahler geometry on finite dimensional complex projective spaces}
\label{sec:Ku}

Consider a finite $N$-dimensional Hilbert space $\Hi$ whose Hermitian product is denoted by $\langle x, x'\rangle_{\Hi}$ and is by convention $\C$-linear with respect to the second entry and anti-linear with respect to the first entry. If $\{e_a\}_{a=1,\dots,N}$ is an Hermitian basis for $(\Hi, \langle~,~\rangle_{\Hi})$, the corresponding coordinates for $x$ are written 
\beq
\label{eq0}
\langle e_a,x\rangle_{\Hi}\,=\,q_a+ip_a
\eeq
with $(q_a,p_a)\,\in\,\R$. The Hilbert space can then be studied as a real $2N$-dimensional manifold $\Hi_{\R}\,\simeq\,\R^{2N}$, with a global coordinate chart given by as above. Upon identifying the tangent space $T_x\Hi_{\R}$ with $\Hi_{\R}$ itself, the Hermitian product acts as
\begin{align}
&\langle \frac{\del}{\del q_a}\,,\,\frac{\del}{\del q_b}\rangle_{\Hi}=\,
\langle \frac{\del}{\del p_a}\,,\,\frac{\del}{\del p_b}\rangle_{\Hi}=\,\delta_{ab}, \nn \\
&
\langle \frac{\del}{\del q_a}\,,\,\frac{\del}{\del p_b}\rangle_{\Hi}=\,
-\langle \frac{\del}{\del p_a}\,,\,\frac{\del}{\del q_b}\rangle_{\Hi}=\,i\delta_{ab}
\label{eq1}
\end{align}
and can then be written as the tensor\footnote{We use the convention that quantities with repeated indices, unless specified, are summed over.} 
\beq
\label{eq2}
h\,=\,\left(\dd q_a\otimes\dd q_a\,+\,\dd p_a\otimes\dd p_a\right)\,+\,i\,\left(\dd q_a\otimes\dd p_a\,-\,\dd p_a\otimes\dd q_a\right)\,=\,\rm g\,+\,i\,\omega
\eeq
whose real component $\rm g$ is the Euclidean metric on $\Hi_{\R}$ while its imaginary component reads the canonical symplectic 2-form $\omega$. The (1,1) tensor on $\Hi_{\R}$  
\beq
\label{eq3}
{\rm J}\,=\,\frac{\del}{\del p_a}\,\otimes\,\dd q_a\,-\,
\frac{\del}{\del q_a}\,\otimes\,\dd p_a,
\eeq
with ${\rm J}^2\,=\,- 1$,  gives the complex structure compatible with both $\mathrm g$ and $\omega$ since 
\begin{align}
&\mathrm g(Ju,v)\,=\,\omega(u,v), \nn \\ &\mathrm g(Ju,Jv)\,=\,\mathrm g(u,v),  \nn \\ & \omega(Ju,Jv)\,=\,\omega(u,v)
\label{eq4}
\end{align}
 for any pair of vector field $u,v$ on $\Hi_{\R}$. These lines allow to directly \cite{koba-nomi} recover $(\Hi_{\R}, \rm J, \rm g, \omega)$   
as a K\"ahler manifold, with torsionless $J$ (a global integrability condition for the complex structure) and closed 2-form $\omega$. 
Moreover, the relations \eqref{eq4} show that $J$ is at the same time both a finite and an infinitesimal generator for transformations preserving the metric and the symplectic structures.

Upon adopting global coordinates $z_a\,=\,q_a+ip_a$ and $\bar z_{a}=q_a-ip_a$ one can write
\begin{align}
&\mathrm g\,=\,\frac{1}{2}\,(\dd z_a\otimes\dd \bar z_a\,+\,\dd \bar z_a\otimes\dd  z_a), \nn \\ &\omega\,=\,\frac{i}{2}\,\dd z_a\wedge\dd \bar z_a, \nn \\  &\mathrm J\,=\,i\,\left(\frac{\del}{\del z_a}\,\otimes\,\dd z_a\,-\,\frac{\del}{\del \bar z_a}\,\otimes\,\dd \bar z_a\right). \label{eq5}
\end{align}
It is easy to see from \eqref{eq2} that the group of linear maps in $\Hi_{\R}$ leaving the Hermitian tensor $h$ invariant (i.e. the unitary group for the given $h$ tensor) is equivalently  given as one of the intersections 
\beq
\label{eq6}
{\rm U}(N)\,=\,{\rm O}(2N, \R)\cap{\rm Sp}(2N,\R)\,=\,{\rm GL}(N,\C)\cap{\rm O}(2N, \R)\,=\,{\rm Sp}(2N,\R)\cap{\rm GL}(N,\C),
\eeq
 where the orthogonal group refers to the real part ${\rm g}$ and the symplectic group refers to the imaginary part ${\omega}$ of $h$. Consider a matrix $W\in\mathbb M^{2N}(\R)$ and the associated linear vector field $X_W\,=\,W_{ab}x_b\del_a$ (where we have collectively denoted the coordinate functions by $\{x_a\}_{a=1,\dots,2N}$). The one parameter group of linear transformations generated by $X_W$ is then unitary on $(\R^{2N},h)$  if one of the following  sets of conditions is fullfilled (by $L_{X_W}$ we denote the Lie derivative of a tensor along $X_W$):
\begin{itemize}
\item it is $L_{X_W}{\rm J}=0$ \emph{and} $L_{X_W}{\rm g}=0$;  
\item it is $L_{X_W}{\rm J}=0$ \emph{and} $L_{X_W}{\omega}=0$;
\item it is $L_{X_W}{\rm g}=0$ \emph{and} $L_{X_W}\omega=0$.
\end{itemize} 
One easily  sees that linear unitary maps on $(\R^{2N},h)$ are infinitesimally generated by vector fields that can be identified with matrices $W$ such that
\beq
\label{eq10}
W\,=\,\begin{pmatrix} A & B \\ -B & A\end{pmatrix} \qquad{\rm with} \quad A=-A^{T}, \,B=B^T
\eeq
These elements define the matrix Lie algebra $\mathfrak u_N$, with ${\rm dim}_{\R}\mathfrak u_N=N^2$. The linear vector fields $X_W$ are  both Hamiltonian and of Killing type: we refer to them as Hermitian vector fields.  This name is natural, since the corresponding Hamiltonian function $f_W$  can be written in terms of a quadratic form associated to $W$, namely\footnote{We shall also use the Dirac's bra-ket notation for elements $|z\rangle\,=\,(z_1,\dots,z_N)$ in $\Hi\simeq\C^N$.}
\beq
\label{eq11}
f_W(q,p)\,=\,\frac{1}{2}\begin{pmatrix} q_a & p_a \end{pmatrix} \,\begin{pmatrix} B_{ab} & -A_{ab} \\ A_{ab} & B_{ab} \end{pmatrix} \,\begin{pmatrix} q_b \\ p_b \end{pmatrix}\,=\,\frac{1}{2}\bar z_a(H_{ab})z_b\,=\,\langle z|H|z\rangle 
\eeq
where $H\in{\mathbb M}^N(\C)$ is given by $H=B+iA=H^{\dagger}$. 
For Hermitian vector fields it is immediate to prove the following identities, which will be useful through the rest of the paper,
\begin{align}
&\omega(X_{H_1}, X_{H_2})\,=\,f_{-i[H_1H_2-H_2H_1]}, \nn \\
&{\rm g}(X_{H_1},X_{H_2})\,=\,f_{(H_1H_2+H_2H_1)}.
\label{eq12} 
\end{align}
The unitary dynamics generated by  $X_W$ on  $\R^{2N}$ 
can be written as
\beq
\label{eq13}
i\frac{\dd z_a}{\dd t}\,=\,H_{ab}z_b, \qquad\qquad 
-i\frac{\dd \bar z_a}{\dd t}\,=\,\bar H_{ab}\bar z_b
\eeq
so we can identify the holomorphic sector of $\C^{2N}$ with $\R^{2N}$ and conclude that the   Schr\"odinger equation \eqref{eq13}   on a finite dimensional Hilbert space $(\Hi, h)$ is given by a Hermitian vector field on the  associated K\"ahler manifold $(\Hi_{\R}, {\rm J}, {\rm g}, \omega)$.  As we mentioned in the introduction, the pairing between the set of states and the set of observables makes the difference between a unitary classical dynamics on the canonical phase space $\R^{2N}$ and a quantum dynamics on $\Hi=\C^N$. We turn our attention to the set of pure states for a finite level quantum dynamics.

\subsection{A reduction procedure}
\label{suse:red}

It is well known \cite{koba-nomi} that, for a finite dimensional $\Hi$,  the projective space $\PP(\Hi)$ has a K\"ahler structure. In order to describe how this can be introduced within a reduction formalism, we start by considering the  example   $\Hi=\C^2$. The projective space is the quotient $\PP(\C^2)=\C^2_0\backslash \C_0$ with respect to the action of $u\in\C_0$ upon $(z_1,z_2)$ given by $(uz_1,uz_2)$ with $z_1\bar z_1+z_2\bar z_2\neq0$. The properties of this action show  that $\PP(\C^2)$ is the basis of the principal bundle $\pi\,:\,\C^2_0\,\stackrel{\C_0}{\longrightarrow}\,\PP(\C^2)$ with fiber given by the (2 dimensional abelian) Lie group $\C_0$. The infinitesimal generators for the action of such a  group provide the vertical fields for the fibration. They are 
\begin{align}
&\Delta \,=\,q_1\dequ+q_2\deqd+p_1\depu+p_2\depd, \nn \\
&\Gamma\,=\,(p_1\dequ+p_2\deqd -q_1\depu -q_2\depd): 
\label{eq7}
\end{align}
the Euler vector field $\Delta$ generates the dilation on $\R_0^4$ associated to the multiplicative $\R_0^+$ subgroup in $\C_0$,  the vector field $\Gamma$ generates the rotation on $\R^4_0$ associated to the ${\rm U}(1)$ subgroup in $\C_{0}$. 
The fibration we are considering is well known. Since the  group $\C_0$ is abelian, we can describe it  as the compositions of a  Kustaanheimo -- Stiefel projection $\pi^{\Delta}$  with a ${\rm U}(1)$ Hopf projection  $\pi^{\Gamma}$ \cite{cimm, damv} equivalently as follows 
\begin{align}
\pi\,&:\,\R^4_0\quad\stackrel{{\rm U}(1)}{\longrightarrow}\quad\R^3_0\quad\stackrel{\R^+_0}{\longrightarrow}\quad{\rm S}^2, \nn \\ &:\,\R^4_0\quad\stackrel{\R^+_0}{\longrightarrow}\quad {\rm S}^3\quad\stackrel{{\rm U}(1)}{\longrightarrow}\quad {\rm S}^2.
\label{eq14}
\end{align}
Since it generates a unitary action on $\R^4$, we have that $\Gamma$ is Hermitian.  Its  Hamiltonian function \eqref{eq11} is given by $H_{ab}=\delta_{ab}$ so we write 
\beq
y_{\gamma}\,=\,\frac{1}{2}(q_1^2+p_1^2+q_2^2+p_2^2)\,=\,\frac{1}{2}r^2.
\label{eq8}
\eeq
We complete the set $\{\Delta, \Gamma\}$ to a system of generators for the space of vector fields on $\C^2_0$ which is suitable for the reduction associated to the fibration we wrote.  We start by noticing that ${\rm g}(\Delta, X_H)=0$ on $\R^{2N}_0$ for any Hermitian vector field $X_H$. 
From the second line in \eqref{eq12} we see also  that a realization of the Clifford algebra for the 3d Euclidean metric in terms of Hermitian matrices on $\C^2$ provides a set of orthogonal Hermitian fields on $\C_0^{2}$. The identification $H_j=\sigma_j$ 
with $\sigma_j$ the Pauli matrices gives the Hermitian vector fields (with the corresponding Hamiltonian functions $y_j$, see \eqref{eq11})
\begin{align}
X_1\,=\,(p_2\dequ+p_1\deqd -q_2\depu -q_1\depd),& \qquad\qquad y_1\,=\,(q_1q_2+p_1p_2) \nn \\
X_2\,=\,(-q_2\dequ+q_1\deqd -p_2\depu +p_1\depd),&\qquad\qquad y_2\,=\,(q_1p_2-q_2p_1) \nn \\
X_3\,=\,(p_1\dequ-p_2\deqd -q_1\depu +q_2\depd), &\qquad\qquad y_3\,=\,\frac{1}{2}(q_1^2+p_1^2-q_2^2-p_2^2),
\label{eq9}
\end{align}
with $y_{\gamma}^2=y_1^2+y_2^2+y_3^2$.  The Hermitian vector fields $X_j$ are the generators of the natural left  action  of the ${\rm SU}(2)$ subgroup of the ${\rm U}(2)$ group on $\R^4_0$, providing a global basis of right invariant vector fields for the tangent space  to the group manifold ${\rm SU}(2)\simeq {\rm S}^3$. The set $\{\Delta, X_j\}$ gives a global orthogonal  basis for the tangent space to $\R^4_0$, with ${\rm g}(X_j,X_k)\,=\,2y_{\gamma}\delta_{jk}$ and clearly ${\rm g}(\Delta, \Delta)=2y_{\gamma}$.

Now we wonder: is it possible to define a suitable reduction procedure that, along  the fibration \eqref{eq14}, allows to induce a K\"ahler structure onto ${\rm S^2}\simeq\mathbb P(\C^2)$ starting from $(\C^2, {\rm J}, {\rm g}, \omega)$? 

We start by recalling that, given a principal bundle $\pi:P\,\stackrel{{\rm G}}{\longrightarrow}\,B$ with gauge group $\rm G$ and vertical fields $V_i\in\mathfrak{X}(P)$, one has that the algebra $\mathcal F(B)$ of functions on the basis of the bundle can be written as the subalgebra $\mathcal F(B)\,=\,\{f\in\mathcal F(P)\,:\,L_{V_i}f=0\}$. The idea to characterize \emph{projectable} vector fields for the fibration is to analyse under which conditions are vector fields $D\in\mathfrak X(P)$ derivations for $\mathcal F(B)$. One can prove that the vector field $D$ is projectable if and only if\footnote{Notice that this notions parallels that of \emph{normaliser} of a subalgebra $V$ of a Lie algebra $\mathfrak X$.} the commutator $[D,V_i]$ is vertical for any vertical $V_i$.  

This notion naturally generalises to the study of the projectability of any contravariant tensor field on the total space of a bundle. We then consider, on $\R^4_0$, the tensors $(a=1,2)$
\begin{align}
&G\,=\,\dequ\otimes\dequ\,+\,\deqd\otimes\deqd\,+\,\depu\otimes\depu\,+\,\depd\otimes\depd\,=\,2(\frac{\del}{\del z_a}\otimes \frac{\del}{\del \bar z_a}\,+\,\frac{\del}{\del \bar z_a}\otimes \frac{\del}{\del z_a}) \nn \\
&\Lambda\,=\,\dequ\wedge\depu\,+\,\deqd\wedge\depd\,=\,2i(\frac{\del}{\del z_a}\wedge\frac{\del}{\del \bar z_a})
\label{eq15}
\end{align}
It is evident that $G$ gives the Euclidean metric on $\R^4$ in contravariant form while $\Lambda$ is the Poisson tensor corresponding to the canonical symplectic structure $\omega$. Both tensors turn out to be projectable with respect to the ${\rm U}(1)$ subgroup action with infinitesimal generator $\Gamma$, but not with respect to the dilation which is infinitesimally generated by the Euler vector field $\Delta$, since their coordinate expressions  are not homogeneous of degree zero  in the linear coordinate chart adapted to $\Delta$.  
We first  notice that
\beq
[X_j, \Delta]\,=\,0, \qquad [X_j,\Gamma]\,=\,0,
\label{eq16}
\eeq
so the vector fields $\{\Delta, \Gamma, X_j\}$ are projectable, with clearly $\pi_*(\Gamma)=0$ and $\pi_*(\Delta)=0$, then observe also that the tensors
\beq
\label{eq17}
\tilde{G}\,=\,(\bar z_1z_1+\bar z_2z_2)\,G, \qquad\qquad \tilde{\Lambda}\,=\,(\bar z_1z_1+\bar z_2z_2)\,\Lambda
\eeq
\emph{are} now projectable, since the factor $(\bar z_1z_1+\bar z_2z_2)=r^2=2y_{\gamma}$ is invariant under the action of $\Gamma$ and both $\tilde{G}$ and $\tilde{\Lambda}$ are  homogeneous of degree $0$. A direct computation moreover reads
\begin{align}
&\tilde{G}\,=\,\Delta\otimes\Delta\,+\,X_j\otimes X_j, \nn \\ 
&y_{\gamma}\tilde{\Lambda}\,=\,\varepsilon_{abc}y_aX_b\wedge X_c\,+\,y_{\gamma}\Gamma\wedge\Delta.
\label{eq18}
\end{align}
The projection along $\Gamma$ has a coordinate expression given by the Hamiltonian functions $\pi^{\Gamma}\,:\,(q_a,p_a)\to (y_{\gamma}, y_j)$. It becomes immediate to compute that  clearly $\pi^{\Gamma}_*(\Gamma)=0$ and 
\begin{align}
&\mathfrak{X}(\R^3_0)\,\ni\,\pi^{\Gamma}_*(\Delta)\,=\,2(y_{\gamma}\frac{\del}{\del y_{\gamma}}\,+\,y_j\frac{\del}{\del y_j})\,=\,\tilde{\Delta}, \nn \\
&\mathfrak{X}(\R^3_0)\,\ni\,\pi^{\Gamma}_*(X_j)\,=\,2\varepsilon_{jab}y_{a}\frac{\del}{\del y_b}\,=\,\tilde{X}_j \label{eq19}
\end{align}
for the projected vector fields, thus recovering $y_j\tilde X_j=0$, with the space of vector fields tangent to ${\rm S}^2$ being  not a free module. We have now to project along 
$\tilde\Delta$, and this  amounts to fix a value for $y_{\gamma}$, i.e. the radius for ${\rm S}^2$ embedded in $\R^3_0$. If we set $r^2=1$, that is $y_{\gamma}=1/2$, then  $\pi^{\tilde\Delta}_*(\tilde\Delta)=0$ and 
\beq
\label{eq20}
\pi_*(X_j)\,=\,(\pi^{\tilde\Delta}_*\circ\pi^{\Gamma}_*)(X_j)=2\varepsilon_{jab}y_{a}\frac{\del}{\del  y_b}\,=\,R_j.
\eeq
We write 
\begin{align}
&\pi_*(\tilde G)\,=\,R_a\otimes R_a, \nn \\
&\pi_*(\tilde \Lambda)\,=\,\varepsilon_{abc}y_aR_b\wedge R_c
\label{eq21}
\end{align}
for the projected tensors \eqref{eq18} and prove that they provide ${\rm S}^2\simeq\mathbb P(\C^2)$ its well known K\"ahler 
structure. We start by considering the covariant form  $\tilde{\rm g}$ and $\tilde{\omega}$ of the contravariant tensors written in \eqref{eq21}. They are given by 
\begin{align}
&\tilde{\rm g}(V_a, V_b)\,=\,\pi_*(\tilde{G})(\alpha_a,\alpha_b), \nn \\
&\tilde{\omega}(D_a,D_b)\,=\,\pi_*(\tilde{\Lambda})(\alpha_a,\alpha_b)
\label{eq22}
\end{align}
where the $\mathcal F({\rm S}^2)$-bimodule map $\mathfrak{G}\,:\,\Omega^1({\rm S}^2)\,\ni\,\alpha\,\mapsto\,V\,\in\,\mathfrak{X}({\rm S}^2)$ 
is defined via the duality $\pi_*(\tilde G)(\alpha, \beta)\,=\,\beta(V)$ for any 1-form $\beta$ and the analogous map $\mathfrak{L}\,:\,\Omega^1({\rm S}^2)\,\ni\,\alpha\,\mapsto\,D\,\in\,\mathfrak{X}({\rm S}^2)$  via  $\pi_*(\tilde \Lambda)(\alpha, \beta)\,=\,\beta(D)$. Their coordinate expression is given by 
\begin{align}
\pi_*(\tilde \Lambda)\,=\,2\varepsilon_{abc}y_a\frac{\del}{\del y_b}\wedge\frac{\del}{\del y_c}& \qquad\qquad\tilde{\omega}\,=\,\frac{1}{2}\varepsilon_{abc}y_a\dd y_b\wedge\dd y_c, \nn \\
\pi_*(\tilde G)\,=\,\frac{\del}{\del y_a}\otimes\frac{\del}{\del y_a}&\qquad\qquad\tilde{\rm g}\,=\,\frac{1}{4}\dd y_a\otimes\dd y_a
\label{eq23}
\end{align}
We remark that $\tilde{\rm g}$ comes as the restriction\footnote{One computes explicitly that $\pi_*^{\Gamma}(\tilde G)=4y_0^2\frac{\del}{\del y_a}\otimes\frac{\del}{\del y_a}\,-\,4y_ay_b\frac{\del}{\del y_a}\otimes\frac{\del}{\del y_a}$, which gives the expression in \eqref{eq23} since $y_a\frac{\del}{\del y_a}$ is zero on elements in $\mathcal F({\rm S}^2)$. Analogously, one computes that $\tilde{\rm g}(\frac{\del}{\del y_a},\frac{\del}{\del y_b})\,=\,y_0^2\dd y_a\otimes\dd y_b-y_ay_b\dd y_a\otimes\dd y_b$ which reads the expression in \eqref{eq23} since $y_a\dd y_a=0$ as a 1-form on ${\rm S}^2$.}  to ${\rm S}^2$ of the Euclidean metric tensor on $\R^3_0$, and coincides with the Fubini-Study metric for $\rm S^2$.
To study the complex structure on ${\rm S}^2$ we notice that, in analogy to \eqref{eq18}, one considers $\tilde{\rm J}=2y_{\gamma}{\rm J}$, with  
\beq
\label{eq24}
\tilde{\rm J}\,=\,\Delta\otimes\theta_{\gamma}\,-\,X_k\otimes\dd y_k
\eeq
where $\theta_{\gamma}\,=\,p_a\dd q_a-q_a\dd p_a$ is the canonical connection 1-form for the Hopf ${\rm U}(1)$ fibration we are considering, while  $(\dd y_k)$ are the differentials of the Hamiltonian functions $y_k$. The relation \eqref{eq24} shows that the operator $\tilde {\rm J}_{\pi}$ with 
\beq
\label{eq25}
\tilde {\rm J}_{\pi}\,:\,R_k\quad\mapsto\quad \frac{\del}{\del y_k}, \quad\mapsto\quad-R_k.    
\eeq
is a (1,1)-tensor field \emph{related}\footnote{Since the tensor $\tilde {\rm J}$ is not contravariant, it cannot be projected along the fibration onto ${\rm S}^2$.} to $\tilde{\rm J}$.   
It is now immediate to prove that $(\tilde{\rm g}, \tilde{\omega}, \tilde{\rm J}_{\pi})$ are compatible, i.e.
\begin{align}
&\tilde{\mathrm g}(\pi_*(\tilde{\rm J})u,v)\,=\,\tilde{\omega}(u,v), \nn \\
&\tilde{\mathrm g}(\pi_*(\tilde{\rm J})u,\pi_*(\tilde{\rm J})v)\,=\,\tilde{\mathrm g}(u,v), \nn \\ & \tilde{\omega}(\pi_*(\tilde{\rm J})u,\pi_*(\tilde{\rm J})v)\,=\,\tilde{\omega}(u,v)
\label{eq26}
\end{align}
for any $u,v\in\mathfrak{X}({\rm S}^2)$. The integrability condition for the corresponding K\"ahler structure is in this example trivially satisfied, with $\dd\tilde{\omega}=0$.

We briefly comment on the form of the tensor $\tilde{\rm J}$. The canonical symplectic form $\omega$ defined on  $\R^{2N}$ allows to define, analogously to the map $\mathfrak{L}$ defined above, a duality $\mathfrak{S}\,:\,\mathfrak{X}(\R^{2N})\,\to\,\Omega^1(\R^{2N})$ by $\omega(X',X)\,=\,(\mathfrak{S}(X))(X')$ for any $X,X'\,\in\,\mathfrak{X}(\R^{2N})$. Its explicit expression reads $\mathfrak{S}(\frac{\del}{\del q_a})\,=\,-\dd p_a$ and $\mathfrak{S}(\frac{\del}{\del p_a})\,=\,\dd q_a$. The complex structure in \eqref{eq3} can be written as 
\beq
\label{eq26a}
{\rm J}\,=\,\frac{\del}{\del p_a}\otimes \mathfrak{S}(\frac{\del}{\del p_a})\,+\,\frac{\del}{\del q_a}\otimes \mathfrak{S}(\frac{\del}{\del q_a})
\eeq
where the set $\{\frac{\del}{\del p_a}, \frac{\del}{\del p_a}\}$ gives an orthonormal basis for the space of derivations on $\R^{2N}$. If we consider the restriction of $\omega$ from $\R^4$ to $\R^4_0$ then we immediately compute that $\mathfrak{S}(\Delta)=\theta_{\gamma}$ while $\mathfrak{S}(X_k)\,=\,-\dd y_k$. We can then clearly write for \eqref{eq24}
\beq
\label{eq36}
\tilde{\rm J}\,=\,\Delta\otimes\mathfrak{S}(\Delta)\,+\,X_k\otimes\mathfrak{S}(X_k).
\eeq
Notice that the difference between $\tilde{\rm J}$ and the restriction of the canonical ${\rm J}$ to $\R^4_0$ comes by the choice of a basis $\{\Delta, X_k\}$ for the space of derivations in $\R^4_0$ which is orthogonal but not orthonormal.

The reduction procedure we described can be generalised to equip any finite dimensional projective space $\mathbb P(\C^N)$ with the Fubini-Study metric and the corresponding K\"ahler structure. The fibration $\pi\,:\,\C^N_0\,\stackrel{\R^+_0}{\longrightarrow}\,{\rm S}^{2N-1}\,\stackrel{{\rm U}(1)}{\longrightarrow}\,\mathbb P(\C^N)$ is along the vertical vector fields $\Delta\,=\,q_a\frac{\del}{\del q_a}+p_a\frac{\del}{\del p_a}$ and $\Gamma\,=\,-{\rm J}(\Delta)\,=\,
-q_a\frac{\del}{\del p_a}+p_a\frac{\del}{\del q_a}$

We conclude this section by noticing also that such a reduction procedure 
is meaningful and provides the correct well known  K\"ahler structure on $\mathbb P(\C^N)$ when applied to the rescaled  tensors $(\tilde{G}=\langle z|z\rangle G, \tilde{\Lambda}=\langle z|z\rangle \Lambda, \tilde{\rm J}=\langle z|z\rangle {\rm J})$ on $\C^N_0$ with $(G,\Lambda, {\rm J})$ coming from the canonical structure as in \eqref{eq2}.  Such tensors \emph{do not} provide $\C^N_0$ a K\"ahler structure: $\tilde\Lambda$ is not a Poisson tensor,  the corresponding bracket $\{f,f'\}=\tilde{\Lambda}(\dd f,\dd f')$  (see \cite{arn-giv, cov-jac}) gives a Jacobi bracket. We close this section by reporting that that this procedure holds true also for an infinite dimensional Hilbert space, see \cite{db}. 

\subsection{Unfolding via the momentum map}
\label{suse:mom}
We have already noticed by the relations \eqref{eq6} and \eqref{eq10} that the Lie algebra $\mathfrak u_N$ of the unitary group ${\rm U}(N)$ is represented by the real vector space of anti-Hermitian matrices, i.e. $\mathfrak{u}_N=\{\mathbb M^N(\C)\,\ni\,T=-T^{\dagger}\}$ with Lie algebra bracket  given by the standard matrix commutator. Since the Cartan-Killing form is not degenerate, we identify the real vector space  $\mathfrak{u}^*_N=\{A=A^{\dagger}\in\mathbb M^N(\C)\}$ of  Hermitian matrices with the dual to $\mathfrak{u}_N$ via the pairing $A(T)\,=\,i{\rm Tr}(AT)/2$. The real vector space isomorphism   defined by $\mathfrak{u}_N^*\,\ni\,A\,\mapsto\, -iA=\hat{A}\,\in\,\mathfrak{u}_N$  allows to define a scalar product in $\mathfrak{u}_N$ via $\langle\hat A,\hat B\rangle_{\mathfrak{u}}=\langle A,B\rangle_{\mfu^*}={\rm Tr}(AB)/2$ and a Lie algebra bracket $[A,B]_{\mfu^*}=[\hat A,\hat B]=-i[A,B]$. 
 The set $\{\sigma_{\alpha}\}_{\alpha=1,\dots,N^2}$ denotes an orthonormal basis for $\mfu^*_N$ with respect to such a scalar product, the elements  $\hat{\sigma}_{\alpha}=-i\sigma_{\alpha}=\tau_{\alpha}$ provide the dual basis in $\mfu_N$. 

The action of the unitary group ${\rm U}(N)$ on $\C^N$ is Hamiltonian with respect to the canonical symplectic structure in \eqref{eq2}, since the infinitesimal generators are the Hamiltonian (see \eqref{eq11}) vector fields $X_{H}$ with $H\in\mfu^*_N$ and  from \eqref{eq12}  the Poisson bracket between the corresponding  Hamiltonian functions is  $\{f_{H_1}, f_{H_2}\}=f_{[H_1,H_2]_{\mfu^*}}$. Such  Hamiltonian action of ${\rm U}(N)$ allows to define a momentum map $\mu\,:\,\C^N\,\to\,\mfu^*_N$ given by $(\mu(z))(\tau_{\alpha})\,=\,f_{\sigma_{\alpha}}$. An immediate computation shows that 
\beq
\label{eq29}
(\mu(z))(\tau_{\alpha})\,=\,\frac{1}{2}{\rm Tr}(\mu(z)\sigma_{\alpha})\,=\,f_{\sigma_{\alpha}}\,=\,\frac{1}{2}\langle z|\sigma_{\alpha}z\rangle
\eeq
so that the momentum map can be written as
\beq
\label{eq27}
\mu(z)\,=\,|z\rangle\langle z|.
\eeq
Its range is the space $\mathcal P^1$ of not negative, Hermitian and rank 1 matrices on $\C^N$. From \eqref{eq29} we see that any element of such a  range can be written as
\beq
\label{eq28}
|z\rangle\langle z|\,=\,f_{\sigma_{\alpha}}(z,\bar z)\,\sigma_{\alpha}
\eeq
so we can consider $\mathcal P^1$ as a real submanifold in $\mfu^*_N$, 
with local coordinate system\footnote{Notice that this generalises what we have considered for the $\C^2$ example in the previous pages.} given by $y_{\alpha}$ with $\mu^*y_{\alpha}=f_{\sigma_{\alpha}}$. Since $\mfu^*_N$ is a finite dimensional vector space, we identify its  tangent and  cotangent space at each point $\rho\in\mfu^*_N$  as $T_{\rho}\mfu^*_N\simeq\mfu^*_N\oplus\mfu^*_N$ and $T^*_{\rho}\mfu^*_N\simeq\mfu^*_N\oplus\mfu_N$, writing down  identifications at each point as $\mfu^*_N\ni A\,=\,A_{\alpha}(y)\sigma_{\alpha}\,\leftrightarrow\,W_A\,=\,A_{\alpha}(y)\frac{\del}{\del y_{\alpha}}\in\mathfrak{X}(\mfu^*_N)$ and  
$\mfu_N\ni \hat A\,=\,\hat A_{\alpha}(y)\tau_{\alpha}\,\leftrightarrow\,\hat A_{\alpha}(y)\dd y_{\alpha} \in\Omega^1(\mfu^*_N)$.
The scalar product in $\mfu^*_N$ is naturally extended to a scalar product in  $\mathfrak{X}(\mfu^*_N)$, with $\langle W_A,W_B\rangle_{\mathfrak{X}(\mfu^*_N)}\,=\,\langle A,B\rangle_{\mfu^*}$, while the duality between vector fields and 1-forms on $\mfu^*_N$ is clearly given by $\hat A(W_B)\,=\,i{\rm Tr}(AB)/2$.

Given the unitary action $Uz=U|z\rangle$ of  $U\,\in\,{\rm U}(N)$ upon $\C^{N}$, one has 
\beq
\label{eq30}
\mu(Uz)\,=\,U|z\rangle\langle z|U^{\dagger},
\eeq
 so it is evident that $\mu(Uz)=\mu(z)$ if and only if $[U,\mu(z)]=0$. 
For any $0\neq z\in\C^N$, we denote by  $\mathcal O_z$ the orbit for the action of the group ${\rm U}(N)$ through $\mu(z)$.
Since any 1-parameter group of unitary transformations is written as  $U(s)=\exp{(-isA)}$ with $A=A^{\dagger}\in\mfu^*_N$, the infinitesimal generator of this action on $\mu(z)$ gives the  vector field  
\beq
\label{eq31}
W\,=\,\frac{1}{2}{\rm Tr}\left([A,\mu(z)]_{\mfu^*}\sigma_{\alpha}\right)\frac{\del}{\del y_{\alpha}}\,\in\,\mathfrak{X}({\mathcal O}_z).
\eeq
It is then clear that there is a bijection between the elements in the  tangent space in $\mu(z)$ to the orbit $\mathcal O_z$ and the set of Hermitian matrices which can be written as $[A,\mu(z)]_{\mfu^*}$ with $A=A^{\dagger}$. 
Select a  basis $\{|z\rangle, |e_a\rangle\}$ for $\C^N$ (with $a=1,\dots,N-1$) whose vectors satisfy the conditions  $\langle e_a|e_b\rangle_{\C^N}=\delta_{ab}$ and  $\langle z|e_a\rangle_{\C^N}=0$.  It is a long but straightforward calculation to prove that the range of the commutator  $[A=A^{\dagger}, \mu(z))]_{\mfu^*}$ is a real $2(N-1)$ dimensional vector space with a basis given by 
\begin{align}
&\phi_a\,=\,|e_a\rangle\langle z|\,+\,|z\rangle\langle e_a|, \nn \\
&\psi_a\,=\,i(|z\rangle\langle e_a|\,-\,|e_a\rangle\langle z|).
\label{eq32}
\end{align}
From the identities (denote  $\langle z|z\rangle_{\C^N}=\|z\|^2$) 
\begin{align}
&[\phi_a,\mu(z)]_{\mfu^*}\,=\,\|z\|^2\psi_a,\nn \\
&[\psi_a,\mu(z)]_{\mfu^*}\,=\,-\|z\|^2\phi_a, 
\label{eq33}
\end{align}
with
\begin{align}
&\langle \phi_a,\psi_b\rangle_{\mfu^*}\,=\,0, \nn \\
&\langle \phi_a,\phi_b\rangle_{\mfu^*}\,=\,\|z\|^2\delta_{ab}, \nn \\ 
&\langle \psi_a,\psi_b\rangle_{\mfu^*}\,=\,\|z\|^2\delta_{ab}
\label{eq34}
\end{align}
and 
\begin{align}
&\hat{\phi}_a([\psi_b,\mu(z)]_{\mfu^*})\,=\,2\,\|z\|^4\delta_{ab}, \nn \\
&\hat{\phi}_a([\phi_b,\mu(z)]_{\mfu^*})\,=\,\hat{\psi}_a([\psi_b,\mu(z)]_{\mfu^*})\,=\,0, \nn \\
&\hat{\psi}_a([\phi_b,\mu(z)]_{\mfu^*})\,=\,2\,\|z\|^4\delta_{ab},
\label{eq35}
\end{align}
we see that $\{\frac{1}{\|z\|}W_{\phi_a},\frac{1}{\|z\|}W_{\psi_a}\}_{a=1,\dots,N}$ gives an orthonormal  basis for the tangent space to the orbit $\mathcal{O}_z$, with dual basis $\{-\frac{1}{2\|z\|}\hat{\phi}_a,
\frac{1}{2\|z\|}\hat{\psi}_a\}_{a=1,\dots,N}$ for the cotangent space.

Consider now the tensors $G,\Lambda$, which give the contravariant form to the Euclidean metric ${\rm g}$ and the symplectic form $\omega$ defined on $\R^{2N}\simeq\C^N$ as in \eqref{eq2}. The action of the push-forward $\mu_*$ of the momentum map allows to define a symmetric contravariant tensor $\mu_*G$ and a bivector field $\mu_*\Lambda$ on $\mfu^*_N$. It is easy to compute that, for any $\hat A,\hat B\,\in\,\Omega^1(\mfu^*_N)$,  
\begin{align}
\mu^*\left((\mu_*G)(\hat A,\hat B)\right)\,=\,f_{AB+BA}, \nn \\
\mu^*\left((\mu_*\Lambda)(\hat A,\hat B)\right)\,=\,f_{[A,B]_{\mfu^*}}
\label{eq37}
\end{align}
When restricted to the cotangent space of the orbit ${\mathcal O}_z$, the tensors $\mu_*G$ and $\mu_*\Lambda$ turn to be non degenerate, with 
\begin{align}
&\mu_*G(\hat\phi_a,\hat\phi_b)\,=\,\|z\|^4\delta_{ab},\nn \\
&\mu_*G(\hat\psi_a,\hat\psi_b)\,=\,\|z\|^4\delta_{ab},\nn \\
&\mu_*G(\hat\phi_a,\hat\psi_b)\,=\,0 
\label{eq38}
\end{align}
and 
\begin{align}
&\mu_*\Lambda(\hat\phi_a,\hat\phi_b)\,=\,0,\nn \\
&\mu_*\Lambda(\hat\psi_a,\hat\psi_b)\,=\,0 ,\nn \\
&\mu_*\Lambda(\hat\phi_a,\hat\psi_b)\,=\,-\|z\|^4\delta_{ab}.
\label{eq39}
\end{align}
If we invert these tensors, as we described in \eqref{eq22}, we have a metric $\tilde{\rm g}$ and a 2-form $\tilde{\omega}$ on $\mfu^*_N$ which are given by 
\begin{align}
&\tilde{\rm g}(W_{\phi_a},W_{\phi_b})\,=\,\delta_{ab}, \nn \\
&\tilde{\rm g}(W_{\psi_a},W_{\psi_b})\,=\,\delta_{ab}, \nn \\
&\tilde{\rm g}(W_{\phi_a},W_{\psi_b})\,=\,0
\label{eq40}
\end{align}
and 
\begin{align}
&\tilde{\omega}(W_{\phi_a},W_{\phi_b})\,=\,0, \nn \\
&\tilde{\omega}(W_{\psi_a},W_{\psi_b})\,=\,0, \nn \\
&\tilde{\omega}(W_{\psi_a},W_{\phi_b})\,=\,\delta_{ab}.
\label{eq41}
\end{align}
If we define the duality map $\mathfrak{S}\,:\,\mathfrak{X}(\mfu^*_N)\to\Omega^1(\mfu^*_N)$ with respect to $\tilde{\omega}$ in analogy to what we described for $\R^{2N}$ and $\C^{2}_0$ in the previous pages, we introduce the tensor 
\beq
\tilde{\rm J}\,=\,W_{\phi_a}\otimes\mathfrak{S}(W_{\phi_a})\,+\,W_{\psi_a}\otimes\mathfrak{S}(W_{\psi_a})\,=\,\|z\|^2\left(W_{\phi_a}\otimes\hat\psi_a\,-\,W_{\psi_a}\otimes\hat\phi_a\right).
\label{eq42}
\eeq
Notice that the tensor  $\tilde{\rm J}$ is not a complex structure, since $\tilde{\rm J}^2=-\|z\|^4 \bb{I}$. 
Fix now $\|z\|=1$, so that the orbit ${\mathcal O}_z$ coincides with the complex projective space $\mathbb P(\C^N)$. The comparison between \eqref{eq40} and \eqref{eq34} shows that the metric $\tilde{\rm g}$ induced on the complex projective via the momentum map starting from the Euclidean metric ${\rm g}$ on $\R^{2N}\simeq\C^N$  coincides with the restriction to the complex projective space of the natural metric on  $\mfu^*_N$.  It is now possible to prove that $\dd\omega=0$, so that $(\mathbb P(\C^N),\tilde{\rm g}, \tilde{\omega}, \tilde{\rm J})$ is a K\"ahler manifold, and the tensor $\tilde{\rm g}$ coincides with the well known Fubini-Study metric.

We consider the 2 dimensional example within this unfolding procedure. The space $\mfu^*_2$ is the real span of $\{\sigma_0=\bb{I}_2,\sigma_k\}$ with $\sigma_k$ the Pauli matrices, the momentum map reads
\beq
\mu(z)\,=\,y_0\sigma_0\,+\,y_k\sigma_k
\label{eq44}
\eeq
where we have denoted  $y_0=y_{\gamma}$ from \eqref{eq8} and $y_k$ as in \eqref{eq9}. 
Given an orthonormal basis $\{e_1,e_2\}$ for $\C^2$ as in \eqref{eq0}, we consider  $z=(z_1e_1+z_2e_2)/\|z\|$ and $e=(z_2e_1-z_1e_2)/\|z\|$ on $\C^2_0$ with $\|z\|^2=(q_1^2+p_1^2+q_2^2+p_2^2)$ so that we have
\begin{align}
&\phi\,=\,|z\rangle\langle e|+|e\rangle\langle z|\,=\,2(y_1\sigma_3\,-\,y_3\sigma_1), \nn \\
&\psi\,=\,i(|z\rangle\langle e|-|e\rangle\langle z|)\,=\,2(y_0\sigma_2\,+\,y_2\sigma_0)
\label{eq43}
\end{align}
and then 
\begin{align}
&W_{\phi}\,=\,2(y_1\frac{\del}{\del y_3}\,-\,y_3\frac{\del}{\del y_1}), \nn \\
&W_{\psi}\,=\,2(y_0\frac{\del}{\del y_2}\,+\,y_2\frac{\del}{\del y_0})
\label{eq45}
\end{align}
A direct inspection allows to identify the unfolding from $\mathcal O_z$ to $\C^2_0$, which is given by 
\begin{align}
&W_{\phi}\,=\,\mu_*\left(-\frac{1}{\|z\|^2}X_2\right), \nn \\
&W_{\psi}\,=\,\mu_*\left(\frac{2}{\|z\|^4}(y_2\Delta-y_3X_1+y_1X_3)\right)
\label{eq46}
\end{align} 
If we fix $\|z\|=1$, from the metric tensor ${\rm g}$ on $\C^2_0$ it is 
\begin{align}
&{\rm g}((-\frac{1}{\|z\|^2}X_2), (-\frac{1}{\|z\|^2}X_2))\,=\,1\,=\,\tilde{\rm g}(W_{\phi},W_{\phi}), \nn \\
&{\rm g}(\frac{2}{\|z\|^4}(y_2\Delta-y_3X_1+y_1X_3),\frac{2}{\|z\|^4}(y_2\Delta-y_3X_1+y_1X_3))\,=\,1\,=\,\tilde{\rm g}(W_{\psi},W_{\psi}), \nn \\
&{\rm g}(\frac{2}{\|z\|^4}(y_2\Delta-y_3X_1+y_1X_3),(-\frac{1}{\|z\|^2}X_2))\,=\,0\,=\,\tilde{\rm g}(W_{\psi},W_{\phi}).
\label{eq47}
\end{align}
From the symplectic structure $\omega$ we have
\beq
\omega(\frac{2}{\|z\|^4}(y_2\Delta-y_3X_1+y_1X_3),(-\frac{1}{\|z\|^2}X_2))\,=\,1\,=\,\tilde\omega(W_{\psi},W_{\phi})
\label{eq48}
\eeq
The compatibility of the metric tensor with the symplectic structure is immediate to recover once we notice that, given the vector fields \eqref{eq7} and \eqref{eq9} on $\C^2_0$, it is ${\rm J}(\Gamma)=\Delta$ and ${\rm J}(X_k)=2(y_k\frac{\del}{\del y_0}+y_0\frac{\del}{\del y_k})$, which means $\tilde{\rm J}(W_{\phi})=-W_{\psi}$ as it is written in \eqref{eq42} for the projective space.

We stress  that the analysis of projective spaces $\mathbb P(\C^N)$ as a coadjoint orbit of the unitary group ${\rm U}(N)$ on the dual of its Lie algebra $\mathfrak u_N^*$ provides an explicit  (albeit \emph{local}) description of the  corresponding set of vector fields and of 1-forms. This allows to introduce  the Hodge - de Rham Laplacian on $\mathbb P(\C^N)$. An evolution of  such local formulation, with a  \emph{global } description of  the differential calculi on ${\rm SU}(N)$ and a suitable quotient  via  the relevant subgroups, would provide a global description of the Laplacians, since $\mathbb P(\C^N)\,\simeq\,{\rm SU}(N)/({\rm SU}(N-1)\times{\rm U}(1))$. 

On a different level, the formalism we outlined allows to study also the geometry of the set of mixed states. In particular, they can represented as  as Hermitian operators and one can associate with them expectation value functions on the space of pure states. It is then possible to consider Markovian evolutions on them according to the GKLS - master equation. 
We refer to the literature (see \cite{team-par} and references therein) for further details and aim to develop this concluding remarks in forthcoming papers.   

\end{document}